\documentclass{jfm}
\usepackage{graphicx,float}
\usepackage{natbib}

\usepackage{color}

\title[Dynamics of Large Scale Flow in Rayleigh-B\'{e}nard convection]{Dynamics of reorientations and reversals of large scale flow in Rayleigh-B\'{e}nard convection}

\author[P. K. Mishra, A. K. De, M. K. Verma, and V. Eswaran]{P.\ns K.\ns M\ls I\ls S\ls H\ls R\ls A$^1$,\ns A.\ns K.\ns D\ls E$^2$,
\ns M.\ns K.\ns V\ls E\ls R\ls M\ls A$^1$\and V.\ns E\ls S\ls W\ls A\ls R\ls A\ls N$^3$}
\affiliation{$^1$Department of Physics, Indian Institute of Technology, Kanpur~208 016, India\\[\affilskip]
$^2$Department of Mechanical Engineering, Indian Institute of Technology Guwahati~781039, India\\[\affilskip]
$^3$Department of Mechanical Engineering, Indian Institute of Technology, Kanpur~208016, India}

\date{\today}
\pagerange{\pageref{firstpage}--\pageref{lastpage}}
\begin{document}
\label{firstpage}

\maketitle

\begin{abstract}
 We present a numerical study of the reversals and reorientations of the large scale circulation (LSC) of convective fluid in a cylindrical container of aspect ratio one.  We take Prandtl number to be 0.7 and Rayleigh numbers in the range from $6\times10^5$ to $3\times10^7$. It is observed that the reversals of the LSC are induced by its reorientation  along the azimuthal direction, which are quantified using the phases of the first Fourier mode  of the vertical velocity measured near the lateral surface in the mid plane.   During a ``complete reversal'', the above phase changes by around $180^0$ leading to reversals of the vertical velocity at all the probes.  On the contrary, the vertical velocity reverses only at some of the probes during a ``partial reversal'' with phase change other than $180^0$.  Numerically we observe rotation-led and cessation-led reorientations, in agreement with earlier  experimental results. The ratio of the amplitude of the second Fourier mode and the first Fourier mode  rises sharply during the cessation-led reorientations. This observation is consistent with the quadrupolar dominant temperature profile observed during the cessations.  We also observe reorientations involving double cessation. 
\end{abstract}


\section{\label{sec1}Introduction}
Turbulent convection is ubiquitous in nature and in many engineering applications. Rayleigh-B\'{e}nard Convection (RBC) in which fluid confined between two plates is heated from below and cooled on the top is an idealized yet an important paradigm to understand convective turbulence. The dynamics of RBC is governed by the two non-dimensional parameters: the Rayleigh number $R=\alpha \Delta T d^3 g/\nu\kappa$ and the Prandtl number $P=\nu/\kappa$, where $d$ is the vertical height of the container, $g$ is the acceleration due to gravity, $\Delta T$ is the temperature difference between the bottom and top plates, and $\alpha$, $\kappa$, and $\nu$ are the thermal heat expansion coefficient, thermal diffusivity, and kinematic viscosity, respectively, of the fluid.  

\cite{krishnamurti:1981} performed experiments on water ($P \simeq 7.0$) and silicon oil ($P \sim 860$) and observed coherent roll structures, also known as ``large scale circulation'' (LSC),  in the turbulent regime.  Subsequently, \cite{castaing:1989} ascertained the existence of LSC in Helium ($P \simeq0.65$-$1.5$) contained in a cylindrical container. They proposed that coherent large scale structures exist statistically only above a certain Rayleigh Number ($R \simeq 10^8$).  They also observed a low frequency peak in the power spectrum of the temperature field.   \cite{xi:2004} studied the onset of large-scale coherent mean flow in RBC using shadowgraph and particle image velocimetry techniques and showed that LSC is a result of the organization of plume motion.  

\cite{cioni:1997} performed RBC experiments on mercury ($0.021 < P < 0.026$)  and placed several thermistors along the azimuth of the cylinder. They deduced the presence of global circulation from the dipolar temperature distribution measured by the probes.  The temperature fluctuations (after subtracting the mean) switched sign randomly in their experiment.  Since the warmer fluid ascends from one side, and the cooler fluid descends from the other side of the apparatus, \cite{cioni:1997}  deduced that the vertical velocity would also exhibit random ``reversals'' in phase with the temperature fluctuations. This feature of convection has been studied extensively using theoretical, experimental, and computational tools~\cite[see reviews by][]{ahlers:2009,kadanoff:2001}.  In this paper we perform computational investigation of the reversal dynamics of  LSC in a cylindrical geometry. 
 
\cite{cioni:1997}  computed the first Fourier mode of the  measured temperature field.  They observed that the amplitude of the first Fourier mode never vanishes, but the phase of the Fourier mode is highly variable.  A phase change of $\pi$  corresponds to the reversal of the flow.  \cite{cioni:1997} also observed a low-frequency peak in the energy spectrum that corresponds to the circulation frequency of the large-scale flow inside the cylinder. Tsuji {\em et al.}\cite{tsuji:2005} performed an experiment with mercury contained in cylindrical container for aspect ratios $1/2$, $1$, and $2$ and found that the low frequency peak is absent for aspect ratio $1/2$.   \cite{niemela:2001}, and \cite{sreenivasan:2002} reported reversals  in their convection experiments on Helium.  \cite{brown:2005} and \cite{brown:2006} performed similar experiments on water ($P =  4.38$) and measured the temperature field in the bulk using probes  placed at $1/4$th, $1/2$ and $3/4$th height of the cylinder. They observed that the plane containing LSC exhibits diffusive and sometimes quick significant drift in the azimuthal direction with the angular change ($|\Delta {\theta}|$) ranging from small values to $\pi$, i.e., $\pi/4 < |\Delta {\theta}| < \pi$. This is called the ``reorientation'' of LSC. The reorientation of the flow can occur in two ways: (a) through rotation of the circulation plane without any major reduction of the circulation strength; (b) through ``cessation'' of the circulation, followed by a restart in a randomly chosen new direction.   \cite{brown:2006}  also computed the amplitude of the first Fourier mode, and found it to be nonzero for rotation-led reorientations and close to zero during cessation-led reorientations.   \cite{xi:2006} measured azimuthal motion of the LSC using particle image velocimetry and  studied reversals of LSC.  They observed ``double cessation'' in their experiments.    In a similar set of  experiments, \cite{xi:2007},  \cite{xi:2008a},  and \cite{xi:2008}  observed strong dependence of the azimuthal reorientations on the aspect ratio of the cylinder. 

\cite{qiu:2001} performed RBC experiments on water and computed temperature correlation functions.  They reported a transition from random chaotic states to a correlated turbulent states at around $R_c  \simeq 5\times10^7$. However, \cite{sano:1989} report strongly correlated large scale flow beyond $R_c \simeq 4\times10^7$ for helium gas ($P=0.7$).

Various models have been proposed to understand the dynamical behaviour of the LSC.  \cite{sreenivasan:2002} and \cite{benzi:2005} proposed a stochastic model in which the reversal of LSC was explained as noise induced switching between two meta-stable states.  \cite{araujo:2005} attempted to explain the irregular cessation and subsequent reversal of the LSC using the force and thermal balance on a single plume modelled by coupled nonlinear equations related to the Lorenz equations.   \cite{Brown2:2007} and \cite{Brown3:2008} proposed a stochastic model with the strength and the azimuthal orientation of the LSC being determined by two stochastic ordinary differential equations; this model appears to explain the experimental observations of cessation and reorientation.   Villermaux\cite{villermaux:1995} proposed a model to explain the low frequency oscillations in the temperature signal in the bulk. His model assumes that the modes of the boundary layers interact through the slow moving large scale circulation. 

There are only a small number of computational studies on the reorientation or reversal of LSC. \cite{stringano:2006} simulated convection in air ($P=0.7$) in a cylinder with an aspect ratio of $1/2$ and observed a single roll breaking into two counter rotating rolls stacks vertically.   Benzi \& Verzicco\cite{verzicco:2008} performed a simulation of the fluid at $R=6\times10^5$  with white noise added to the heat equation and studied the statistical behaviour of the observed reversals. \cite{breuer:2009} studied RBC for infinite Prandtl number in a 2D box and observed reversals of LSC for very high $R$ ($R=10^9$).   There are others 2D numerical experiments in a box  that report that the reversals of the LSC is due to cessations~\cite[see][]{hansen:1990,hansen:1992} or due to the chaotic movement of rolls perpendicular to the roll axis~\cite[see][]{paul:2010}.

In this paper we study the dynamics of reversals of convective structures using direct numerical simulation (DNS) of turbulent RBC for $P=0.7$  in a cylinder of aspect ratio one. We apply conducting boundary conditions on the top and bottom plates and adiabatic boundary conditions on the lateral walls of the container. No-slip boundary condition is applied on all the surfaces of the container.   Our simulations indicate a LSC in the container.  Furthermore, we observe rotation-led and  cessation-led reorientations of  LSC similar to those observed in experiments.  We compute the  amplitudes and phases of the Fourier modes of the vertical velocity measured near the lateral wall in the mid plane, and establish that the dynamics of the LSC can be captured quite well by the low wavenumber  Fourier modes.

The outline of the paper is as follows. We present our numerical method in Section $2$. The results pertaining to observation of LSC and their reorientations are presented in Section $3$. We conclude our results in Section $4$.

\section{\label{sec2} Numerical Method}

We numerically simulate the convective flow in a cylinder.   The Boussinesq approximation is assumed for the buoyancy in the fluid. The relevant non-dimesionalized dynamical equations for the fluid are   
\begin{eqnarray}
\frac{\partial{\textbf{u}}}{\partial{t}}+ (\textbf{u}\cdot \nabla)\textbf{u} & = & -\nabla\sigma + T \hat{z} 
+ \sqrt{\frac{P}{R}} \nabla^{2}\textbf{u},   \label{eq:u}   \\
\frac{\partial{T}}{\partial{t}}+(\textbf{u}\cdot\nabla)T & = & \frac{1}{\sqrt{PR}}  \nabla^{2}T, \label{eq:T}  \\
\nabla\cdot \bf{u} = 0 \label{eq:incompressible}
\end{eqnarray}
where $\textbf{u}=(u_x,u_y,u_z)$  is the velocity field, $T$ is the temperature field, $\sigma$ is the deviation of pressure from the conduction state, $R=\alpha g (\Delta T) d^3/\nu \kappa$ is the Rayleigh number, $P=\nu/\kappa$ is the Prandtl number, and $\hat{z}$ is the buoyancy direction. Here $\nu$ and $\kappa$ are the kinematic viscosity and thermal diffusivity respectively, $d$ is the vertical height of the container, and $\Delta T$ is the temperature difference between the bottom and top plates.  For the non-dimensionalization we have used $d$ as the length scale, $\sqrt{\alpha (\Delta T) g d}$ (free fall velocity) as the velocity scale, and $\Delta T$ as the temperature scale.  Consequently, $d/\sqrt{\alpha g\Delta T d}$ is the time scale of our simulation.
The aspect ratio of the container is taken to be one.  We confine our study to $P=0.7$ which is a typical Prandtl number for air. 

The above non-dimensionalized equations (\ref{eq:u}-\ref{eq:incompressible}) are solved numerically for a cylindrical geometry using a finite-difference scheme.  The convective parts of the equations are discretized in cylindrical coordinates using Tam and Webb's fourth order central explicit scheme with enhanced spectral resolution~\cite[see][]{tam:1993}. The diffusive part is discretized using the second-order central-difference scheme. For the time advancement, we use the second-order Adam-Bashforth  scheme for the nonlinear terms, and the Crank-Nicholson scheme for the diffusive terms.   We perform simulations for Rayleigh numbers $R=6\times10^5$, $8\times10^6$, $2\times10^7$, and $3\times10^7$. Comparison with experimental results show that these Rayleigh numbers are near the threshold of strong turbulence regimes~\cite[see][]{sano:1989}.

\begin{table}
 
\begin{tabular}{ c c c c c c c c c c c } 
 $ R $&  $N_{r}\times N_{\theta}\times N_{z}$ & $\Delta_{min.}$ & $\Delta_{max.}$ & $l_{max.}$& $\eta_h$ & $N_{BL}$ &$dt$ &$Nu$&$\langle\epsilon^u_c\rangle/\langle\epsilon^u_a\rangle$&$\langle\epsilon^{T}_c\rangle/\langle\epsilon^{T}_a\rangle$\\
& & & & & & & & (comp)& &\\ 
$6\times10^{5}$& $33\times 49 \times 97$ & 0.007 & 0.02   & 0.064 &0.057 & 9 & 0.001 &7.6&0.93& 1.26\\
$6\times10^{5}$& $33\times 90 \times 97$ & 0.006 &0.017 & 0.035  & 0.057 & 9 &0.001 &7.4  &0.93&1.34\\ 
$8\times10^{6}$& $75\times 96 \times 145$ & 0.003 & 0.011 &0.033  & 0.025 & 7 &0.001 &15.1&1.00& 1.33 \\
$8\times10^{6}$&$100\times 120 \times 201$ & 0.002 & 0.008 &0.026 &  0.025 & 9 & 0.0005&15.6&1.10&1.34  \\
$2\times10^{7}$ & $100\times 120 \times 201$ & 0.002 & 0.008 &0.026& 0.018 & 7 & 0.0005&22.1&1.12 &1.35 \\
$2\times10^{7}$ & $100\times 180 \times 201$ & 0.002& 0.007 &0.017& 0.018& 7 & 0.0005&22.3&1.08&1.34 \\
$3\times10^{7}$ & $100\times 120 \times 201$ & 0.0018 & 0.008& 0.026& 0.016 &6 & 0.0005 &24.03&-&- \\ 
\end{tabular}
\caption{$N_r$, $N_\theta$, and $N_z$ are the number of grids along the radial, azimuthal, and vertical directions of the cylindrical container; $\Delta_{min}$ and $\Delta_{max}$ are the minimum and maximum mean-grid sizes; $l_{max} = max(2 \pi r /N_{\theta})$; $\eta_{h}$ is the Kolmogorov length scale calculated using the expression  $\eta_h\simeq\pi(P^2/R * Nu)^{1/4}$;  $N_{BL}$ is the number of points inside the thermal boundary layer; $Nu$(comp) is the Nusselt number obtained from the simulation; $\langle\epsilon^u_c\rangle$ ($=\nu \langle |\nabla {\bf u}|^2 \rangle$) and $\langle \epsilon^T_c \rangle$ ($=\kappa \langle | \nabla T|^2  \rangle $) are the numerically calculated viscous and thermal dissipation rates respectively; $\langle\epsilon^u_a\rangle$ ($=\nu^3(Nu-1)R P^{-2}/d^4$) and $\langle\epsilon^{T}_a\rangle$ ($=\kappa(\Delta T)^2 Nu/d^2$) are  the analytical values of viscous and thermal dissipation rates respectively. The last two columns show $\langle\epsilon^u_c\rangle/\langle\epsilon^u_a\rangle$
and $\langle\epsilon^{T}_c\rangle/\langle\epsilon^{T}_a\rangle$.} 
\label{num_details}
\end{table}

The cylinder volume is discretized into variable grids with finer resolution near the boundary layers. Since the boundary layers significantly affect the dynamics of LSC in convective turbulence, it is necessary to resolve the regions near the top and bottom plates, and the lateral walls~\cite[see][]{stevens:2010}.   In our simulations we choose uniform grids along the azimuthal direction, and non-uniform ones along the radial and vertical directions.   For the grid spacing, the Grotzbach condition~\cite[]{grotzbach:1983} is used, according to which the mean grid size $\Delta~=~(r\Delta\theta \Delta r \Delta z)^{1/3}$ should be smaller than the Kolmogorov and thermal diffusion length scales.   Note that the Kolmogorov length scale ($\eta_h$) is estimated using the formula $\eta_h~\simeq~\pi (P^2/(R * Nu))^{1/4}$, where $Nu$ is the Nusselt number. We also calculate the width of the thermal boundary layer using the formula $\delta_T\sim \frac{1}{2 Nu}$ and ensure that the number of grid points inside the boundary layer ($N_{BL}$) should be greater than 3 to 5. Number of grid points inside the thermal boundary layer is given in Table~\ref{num_details}, and they satisfy the above condition.   

 Our simulations satisfy the Grotzbach condition as the minimum of the mean grid size ($\Delta_{min}$) is smaller than  $\eta_h$ (see Table~\ref{num_details}).  Note however that the maximum grid size $l_{max} = max(2 \pi r /N_{\theta})$ is sometimes larger than $\eta_h$.  To validate our code, we compare the numerically computed Nusselt numbers at two different grids (one more resolved than the other) and find that the Nusselt numbers for the two different grids are quite close.  We also calculate the kinetic energy dissipation rate $\langle \epsilon_c^u \rangle$ ($=\nu \langle |\nabla {\bf u}|^2 \rangle$) 
 and the thermal dissipation rate $\langle \epsilon_c^T \rangle$ ($=\kappa \langle | \nabla T|^2  \rangle $) using the numerical data and compare them with their theoretical estimates $\langle \epsilon_a^u \rangle$ (=$\nu^3(Nu-1)R P^{-2}/d^4$) and $\langle \epsilon_a^T \rangle$ (=$\kappa(\Delta T)^2 Nu/d^2$)~\cite[]{shraiman:1990}. The ratios  $\langle\epsilon^u_c\rangle/\langle\epsilon^u_a\rangle$ and $\langle\epsilon^{T}_c\rangle/\langle\epsilon^{T}_a\rangle$ are listed in the Table~\ref{num_details}. Clearly the numerical kinetic dissipation rate is quite close to its analytical counterpart, with the maximum difference at $12\%$.    For the thermal dissipation, the numerical value always seems to be higher than the analytical value, with the maximum difference around $35\%$.   However note that the theoretical estimates of $\langle \epsilon_c^u \rangle$ and $\langle \epsilon_c^T \rangle$ are functions of the Nusselt number for which we substitute its numerically computed value.     Our numerical results on dissipation rates appear to be less accurate compared to those by~\cite{stevens:2010}.   Also, our numerical thermal dissipation rates are always larger than the theoretical estimate in contrast to those by ~\cite{stevens:2010} who report lower values compared to the corresponding theoretical estimate. The above discrepancies are probably due to the  difference in the time integration schemes and the spatial derivative schemes used by us and~\cite{stevens:2010}.

We use fixed time-step $dt$ which is listed in Table~\ref{num_details}.   Due to the computational complexity, the longest computer run for $R=2\times10^7$ took approximately 45 days  on 32 cores of CHAOS cluster of IIT Kanpur. To ensure that the initial conditions do not affect our final results, we start every run with a conductive state modulated with random noise. We validate our numerical code by comparing our results with the past numerical results.  For example, our numerical results of Nusselt number $Nu$ for different Rayleigh numbers fit with a relation $Nu~\simeq~0.143R^{0.297}$ which is in good agreement with the earlier numerical results of \cite{stringano:2006} and \cite{verzicco:1999},  and the experimental observations of \cite{niemela:2000}. 
\begin{figure}
\includegraphics[angle=270,height=!,width=15cm]{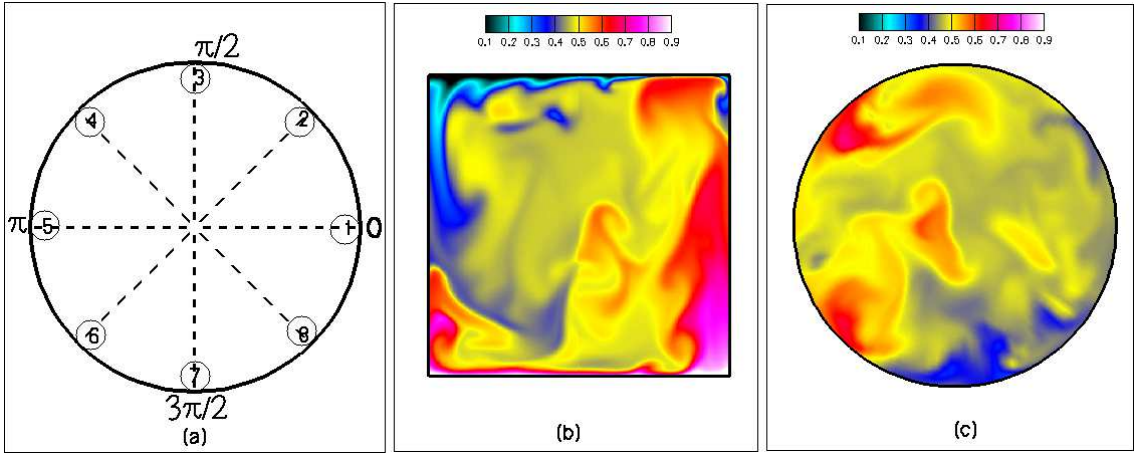}
\caption{(a) Some of the velocity and temperature probes placed inside the cylinder.  For most of our discussions in the paper we use the probes at $z=0.5, r=0.48$ and $\theta= j \pi/4$ with $j=0:7$ shown in the figure. (b)  Temperature profile in a vertical plane $\theta=3\pi/4$ for $R=2\times10^7$ on a ($100\times120\times201$) grid. A hot plume ascends from the right wall and a cold plume descends from the left wall confirming the presence of large scale structure. (c) Temperature profile in a horizontal section $z=0.5$  for $R=2\times10^7$ on a ($100\times120\times201$) grid. }
  \label{schematic_profile}
  \end{figure}
Similar to some of the earlier experiments~\cite[see][]{cioni:1997,brown:2005,brown:2006,xi:2008}, we place ``probes'' in the bulk and in the boundary layers of the cylinder to record the time series of the velocity and temperature fields.  The probes are located at $\theta = \theta_j = j \pi/4$ ($j=0:7$), the vertical heights of $z=0.02, 0.25, 0.5, 0.75, 0.98$, and the radial distances of $r= 0.15, 0.3, 0.48$ from the axis.  The bottom and top probes are inside the thermal boundary layers. A schematic diagram of the probes in the middle plane is depicted in Fig.~\ref{schematic_profile}(a).   The nondimensional  rms values of the velocity $u_{z}^{\rm rms}$, and the eddy turn-over time  ($T_{\rm eddy} = 2d/u_{z}^{\rm rms}$)  for various $R$'s are listed in Table~\ref{num_results}. $u_{z}^{\rm rms}$ has been computed by taking the time average of $\frac{1}{8} (\sum_{i=1}^8(u^{i}_{z})^2)^{1/2}$ for the eight probes placed at $z=0.5$ and $r=0.48$. The eddy turnover time ranges from 26 to 10 free fall times for the range of Rayleigh numbers studied in the present paper; its value decreases with the increase of Rayleigh number consistent with the earlier results of~\cite{qiu:2001}.

\begin{table}
 \begin{center}
\begin{tabular}{ c  c  c  c } 
 $ R $&  $N_{r}\times N_{\theta}\times N_{z}$ & $u_{z}^{\rm rms}$ & $T_{\rm eddy}$\\ 
$6\times10^{5}$& $33\times 49 \times 97$ &0.078&26 \\
$6\times10^{5}$& $33\times 90 \times 97$ &0.083  &24  \\ 
$8\times10^{6}$& $75\times 96 \times 145$ &0.118&17 \\
$8\times10^{6}$&$100\times 120 \times 201$ &0.107&19\\
$2\times10^{7}$ & $100\times 120 \times 201$ &0.185&11 \\
$2\times10^{7}$ & $100\times 180 \times 201$ &0.174&12 \\
$3\times10^{7}$ & $100\times 120 \times 201$ &0.201&10 \\ 
\end{tabular}
\end{center}
\caption{$N_r$, $N_\theta$, and $N_z$ are the number of grids along the radial, azimuthal, and vertical directions of the cylindrical container; $u_{z}^{\rm rms}$ is the rms value of the vertical speed of the flow; and $T_{\rm eddy} =  2d/u_{z}^{\rm rms}$ is the circulation time of large scale flow.} 
\label{num_results}
\end{table}

In the next section we will study the properties of the temperature and velocity time series, and relate them to the earlier experimental results on reversals of LSC.

\section{\label{sec3} Numerical results}

\subsection{Large scale circulation}

The time series measured by the probes of Fig.~\ref{schematic_profile}(a) carries signature of the LSC that will be described below.  Later in the section we will relate the reversals of the vertical velocity field with the reorientation of the LSC.  In Figs.~\ref{schematic_profile}(b) and~\ref{schematic_profile}(c) we display the temperature profiles for $R=2\times10^7$ for a vertical plane ($\theta = 3\pi/4 $) and the middle horizontal plane  ($z=0.5$) respectively. As a supplementary material, we also provide two short movies depicting the flow behaviour for the above two sections.  These figures and movies clearly indicate the presence of convective structures in the flow.  
\begin{figure}
  \begin{center}
  \includegraphics[width=0.78\columnwidth]{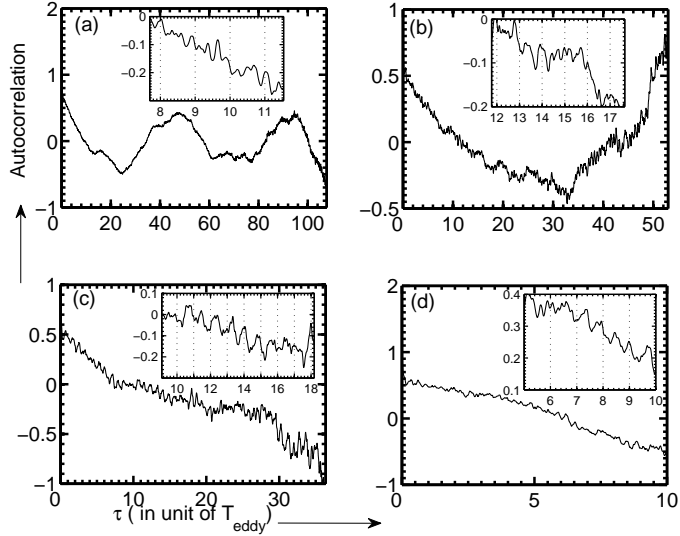}
  \end{center}
  \caption{Plot of the autocorrelation function $g_c(\tau) = \langle \delta T(t) \delta T(t+\tau) \rangle/\sigma_{T}^{2}$  vs. delay time $\tau$ (in units of large eddy turnover time) for a probe at $r=0.48$, $\theta=0$, and $z=0.5$:  (a)$R=6 \times 10^5$ on a ($33\times49\times97$) grid, (b)$R=8\times10^6$ on a ($75\times96\times145$) grid, (c)$R=2\times10^7$ on a ($100\times120\times201$) grid, and (d)$R=3\times10^7$ on a ($100\times120\times201$) grid.  The insets show oscillations on eddy turnover time scale.}
  \label{fig:autocorr}
  \end{figure}

\begin{figure}
  \begin{center}
  \includegraphics[width=0.78\columnwidth]{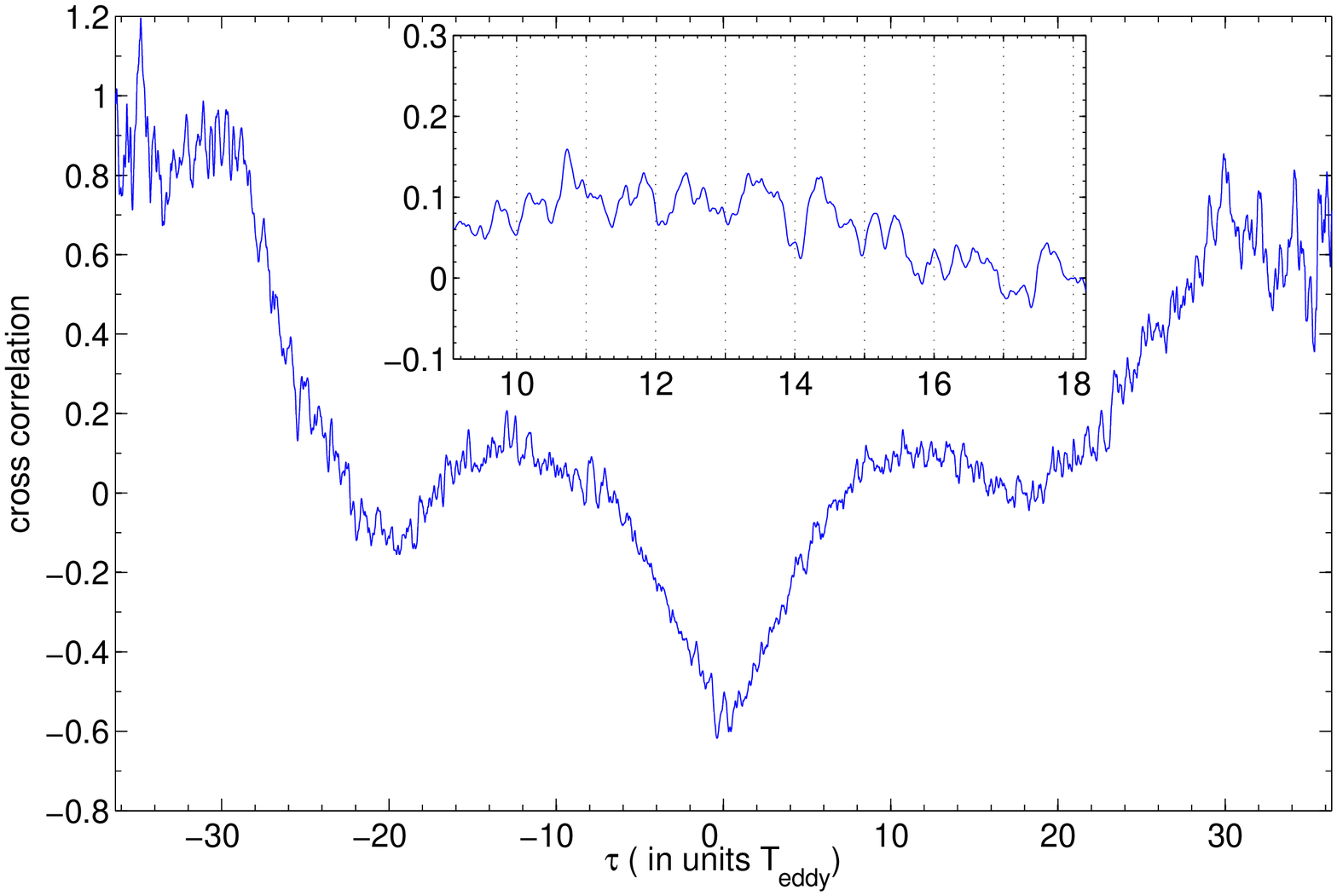}
  \end{center}
  \caption{For $R=2 \times 10^7$ on a ($100\times120\times201$) grid, the cross-correlation function between the temperature measured at two azimuthally opposite probes at $\theta=0$ and $\theta=\pi$ in the mid plane at $r=0.48$. The inset shows oscillations on eddy turnover time scale.}
  \label{fig:cross_cor_ra2e7}
  \end{figure}

In Figs.~\ref{fig:autocorr}(a,b,c,d) we plot the normalized autocorrelation $g_c(\tau) = \langle \delta T(t) \delta T(t+\tau) \rangle/\sigma_{T}^{2}$ at $\theta = 0, r= 0.48, z = 0.5$ for $R=6\times10^5, 8\times10^6, 2\times10^7$, and $3\times10^7$ respectively.  Here  $\delta T(t) = T(t)-\bar{T}$ ($\bar{T}$ is the mean temperature) and $\sigma_{T}^{2}~=~\overline{T(t)^{2}}-(\bar{T})^2$, and time is measured in the units of eddy turnover time.   The autocorrelation functions indicate two time-scales in the system.  Variations at the shorter time-scale, shown in the insets, are due to the statistical return of the convective flow after one eddy turnover time or less (see Table~\ref{num_details}).   For $R=6 \times 10^5$ and $8 \times 10^6$, the oscillations are somewhat irregular.   However for $R=2 \times 10^7$ and $3\times 10^7$,  the oscillations in the insets are quite regular, and the oscillation time period of the auto-correlation function is around one eddy turnover time.   This is also evident from the plot of the cross-correlation  function of temperature between two azimuthally opposite probes placed at $\theta=0$ and $\theta=\pi$, $r=0.48$, and $z=0.5$ (see Fig.~\ref{fig:cross_cor_ra2e7}).  The cross-correlation function also shows oscillations with approximately one eddy turnover time as the time period, in general agreement with the observations of \cite{castaing:1989}, \cite{sano:1989}, \cite{qiu:2001},  and \cite{xi:2006}.  The above correlations are related to the peak in the power spectrum at frequency corresponding to $1/T_{eddy}$.  The peak in the spectrum however tends to be overshadowed by noise.  Consequently the correlation function appears to be a good tool to analyze these oscillations. Note that the autocorrelation studies tend to become strongly periodic when LSC direction is locked, e.g., by a small tilt of the cylinder~\cite[]{Brown:Jstat}.

In addition to the above, we observe variations in the above auto-correlation and cross-correlation functions at much larger time-scales.     For $R=6 \times 10^5$ (see Fig.~\ref{fig:autocorr}(a)), there is a significant decrease in correlations till $\tau \simeq 23 T_{eddy}$, after which correlations rise again to reach a maximum value at $\tau \simeq 56 T_{eddy} $.  These variations possibly correspond to the reorientations of the LSC.  Similar features are observed for  $R=8 \times 10^6$ (see Fig.~\ref{fig:autocorr}(b)).  For $R=2 \times 10^7$ and $3 \times 10^7$, the auto-correlation functions decrease with time (see Figs.~\ref{fig:autocorr}(c,d)).  Incidentally, the cross-correlation function for $R=2 \times 10^7$ shows minima at $\tau =0$ and then at $\tau \approx 20 T_{eddy}$, and a maximum at $ \tau \approx 10 T_{eddy}$, which  possibly indicates reversals of the flow at $ \tau \approx 10 T_{eddy}$, and reoccurrence at $ \tau \approx 20 T_{eddy}$.   We could not perform longer statistics since the computer simulations of RBC flows for large Rayleigh numbers are computationally very expensive.  The auto-correlation and cross-correlation function studies are inconclusive due to averaging, yet, the above observations indicate with some certainty that the LSC exist in RBC.    The reversal time could not be deduced from the present correlation functions.
The presence of LSC becomes more apparent  when we will study the time series of the velocity and temperature fields, a topic of next subsection.

\subsection{Measure of reorientations of LSC}
 \begin{figure}
  \begin{center}
  \includegraphics[height=!,width=15cm]{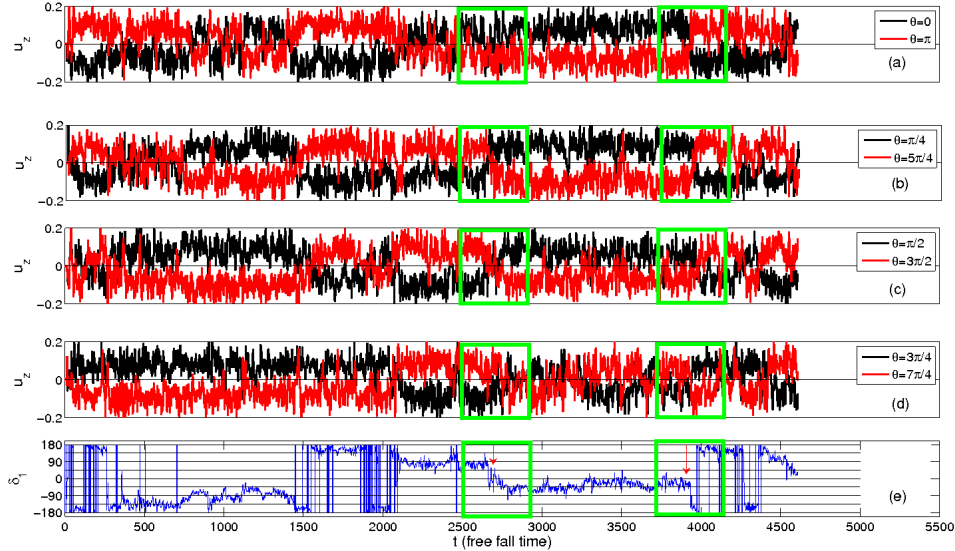}
  \end{center}
\caption{ For $R=6\times10^{5}$ on a ($33\times49\times97$) grid,  the time series of the vertical velocity measured by the probes at $z=0.5$  (mid plane), $r=0.48$: (a) $\theta=0$ and $\theta=\pi$, (b) $\theta=\pi/4$ and $\theta=5\pi/4$, (c) $\theta=\pi/2$ and $\theta=3\pi/2$, and (d) $\theta=3\pi/4$ and $\theta=7\pi/4$. The time series of the phase of the first Fourier mode of the vertical velocity ($\delta_1$) is shown in subfigure (e). Time is measured in the units of the free fall time $d/\sqrt{\alpha g\Delta T d}$. 
The arrows (inside boxed region) in (e) indicate  a partial reorientation ($\delta_1 \simeq 135$ degree)  near $t \simeq 2700$, and  a complete reversal ($\delta_1=180$) near $t \simeq 3900$.} 
  \label{timeseries_r6e5}
  \end{figure}

\begin{figure}
 \begin{center}
  \includegraphics[height=!,width=15cm]{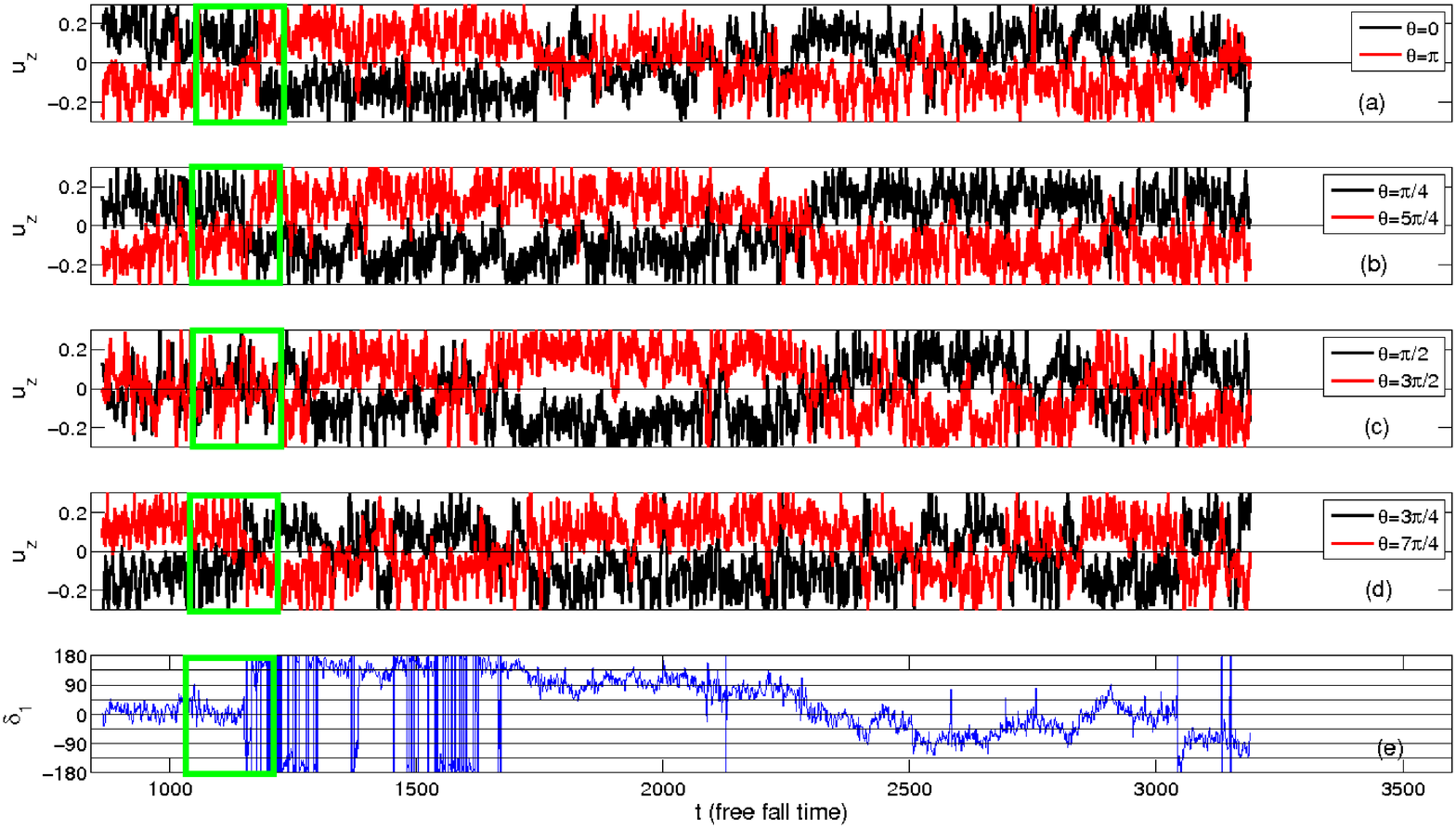}
  \end{center}
   \caption{For $R=8\times10^{6}$ on a ($75\times96\times145$) grid, the time series of the vertical velocity measured by the probes at  $z=0.5$  (mid plane) and $r=0.48$, and the phase of the first Fourier mode of the vertical velocity.  Details of the figures are same as Fig.~\ref{timeseries_r6e5}.  A complete reversal of the flow occurs in the boxed region.}
  \label{timeseries_r8e6}
  \end{figure}

\begin{figure}
  \begin{center}
  \includegraphics[height=!,width=15cm]{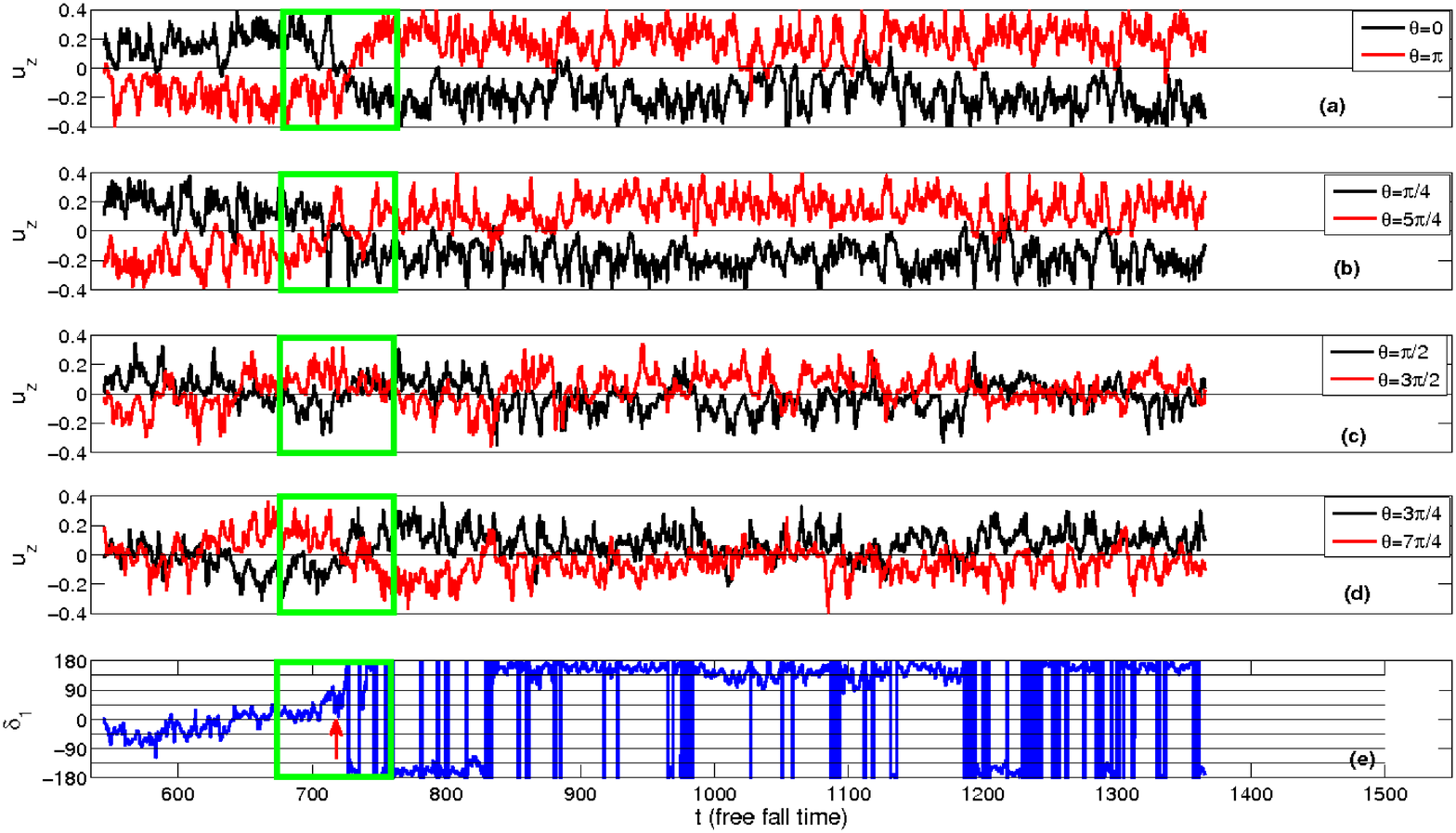}
  \end{center}
  \caption{For $R=2\times10^{7}$ on a ($100\times120\times201$) grid, the time series of the vertical velocity measured by the probes at  $z=0.5$  (mid plane) and $r=0.48$, and the phase of the first Fourier mode of the vertical velocity.  Details of the figures are same as Fig.~\ref{timeseries_r6e5}.   A complete reversal of the flow occurs in the boxed region. }
  \label{timeseries_r2e7}
  \end{figure}

 The time series of the vertical velocity field for $R=6\times 10^5, 8\times10^6$, $2\times10^7 $ recorded by the probes at $\theta_j = j \pi/4$ $ (j=0:7)$, $r=0.48$, and $z=0.5$ are shown in Figs.~\ref{timeseries_r6e5}, \ref{timeseries_r8e6}, and \ref{timeseries_r2e7} respectively. Here time is measured in the units of the free fall time  ($d/\sqrt{\alpha g(\Delta T) d}$).   The figures clearly show that vertical velocity at probes $\theta_j$ and 
 $(\theta_j+\pi) \bmod{2\pi}$  are clearly anti-correlated.   These observations indicate the presence of a large scale convective structure (see Figs.~\ref{schematic_profile}(b,c)).   As shown in Figs.~\ref{timeseries_r6e5}-\ref{timeseries_r2e7}  the local mean value (in time) of the vertical velocity changes sign, and this feature is called ``reversal'' of LSC in literature.  

In the following discussions we will show that the above mentioned reversals are connected to the ``reorientations'' of LSC.  To quantify the reorientations, we Fourier transform the vertical velocity field measured at $r=0.48$ and $z=0.5$.  Note that these probes are near the lateral wall in the mid plane.   The velocity signal at a given probe can be expressed in terms of its Fourier transform as
\begin{eqnarray}
u_j(t)  & = & \sum_{k=-4}^{4} \hat{u}_k \exp{(i k  \theta_j)} \\
          & = & \sum_{k=-4}^{4} |\hat{u}_k|  \exp{(i k  \theta_j + \delta_k)} \\   
          & = & u_{\rm mean} + \sum_{k=1}^{k=4} 2 |\hat{u}_k|\cos{ (k \theta_j + \delta_k)},
\label{eq:fourier}    
\end{eqnarray}
where $\theta_j = j \pi/4$($j=0:7$), and $\delta_k$ is the phase of the $k$-th Fourier mode.  Note  that the reality condition $\hat{u}_{-k} = \hat{u}_{k}^*$ is used to derive Eq.~(\ref{eq:fourier}).   In our simulations we observe that the first Fourier mode ($k=1$) dominates over the other modes except possibly during the reversals.  Note that \cite{cioni:1997}, and \cite{brown:2006} have analyzed the properties of $\hat{u}_1(t)$ and $\delta_1(t)$ only using their experimental data.   In later part of our discussion we will compute the amplitudes of the Fourier modes using the vertical velocity fields measured by probes at $z=0.5$ and $r=0.48$.

For quantitative analysis, \cite{brown:2005}  and \cite{brown:2006} proposed that the phase of the first Fourier mode $\delta_1$ can be used as an approximate measure for the orientation of the plane of LSC.  Using this criteria, they defined ``reorientation of the LSC" as a sudden and significant rotation of the above plane.  They used two selection criteria for the reorientation of LSC: (a) the magnitude of angular change in orientation  ($|\Delta \delta_1|$) should be greater $\pi/4$; (b) the  azimuthal rotation rate ($|\dot{\delta_1}|$) should be greater than $\pi/(5 T_{eddy})$, where $T_{eddy}$ is the eddy turn over time.  We follow the same criteria for the selection of reorientation events in our simulation.

\begin{figure}
  \begin{center}
\includegraphics[width=0.78\columnwidth]{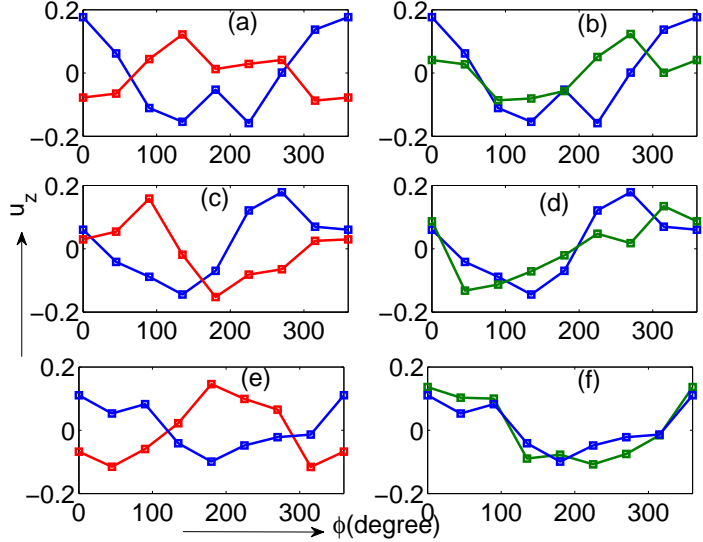}  
  \end{center}
  \caption{Azimuthal profile of the vertical velocity at $z=0.5$ and $r=0.48$ before (blue curve) and after the reorientation (red curve): (a) $R=2\times10^7$ at $t=695$ and 750;  (c)$R = 6\times 10^5$ at $t=2548$ and 2850; (e) $R=6\times10^5$ at $t=3400$ and 4050.   During the reorientations  the phase of the first Fourier mode changes by $\Delta \delta_1 =180$, $135$ and $180$ degrees respectively.  For the reoriented time series we subtract $\Delta \delta_1$ from the positive $k$ modes, and add $\Delta \delta_1$ to the negative $k$ modes, and construct a modified velocity profile (green curve).   As seen in the subfigures (b), (d), and (f), the reconstructed velocity profiles match quite well with the profiles before the reorientation. }
  \label{profile_r6e5}
  \end{figure}

In Figs.~\ref{timeseries_r6e5}-\ref{timeseries_r2e7}  we plot the phase of the first Fourier mode ($\delta_1$), which is a measure of the plane of LSC.  Note that the discontinuities  from $180$ degree to $-180$ degree in the $\delta_1$ diagrams  are not reorientations; they simply indicate jitters near 180 degree.    In all the time series of the vertical velocity and the phase of the first Fourier mode, we observe that the mean value of the vertical velocity changes sign but not necessarily simultaneously for all the probes.  For example in Fig.~\ref{timeseries_r6e5}, the vertical velocity $u_z$ measured by all the probes reverse sign near $t \simeq 3900$ (the right boxed region of Fig.~\ref{timeseries_r6e5}).  This reorientation is ``complete reversal'' of LSC, and it corresponds to the change in $\delta_1$ by around $\pi$, i.e., $\Delta \delta_1 \approx \pi$.
Near $t \simeq 2700$  (the left boxed region of Fig.~\ref{timeseries_r6e5}) however $u_z$ changes sign for all the probes except at $\theta= 0$ and $\pi$.  This kind of reorientation will be termed as a ``partial reversal'', and it corresponds to $\Delta \delta_1 \neq \pi$.  Near $t \simeq 2700$, $\Delta \delta_1 \approx 135^0$.  The time series shown in Figs.~\ref{timeseries_r6e5}-\ref{timeseries_r2e7} exhibit several partial and complete reversals.  For example, we observe complete reversals for $R= 8 \times 10^6$ and $2\times 10^7$ in the boxed regions shown in the  figures. 
A careful analysis of the Fourier modes reveals that the complete and partial reversals are intimately related  to the change in orientations of the convective structures by $\Delta \delta_1$. We take the real space $u_z$ data before and after the reversals for the three cases marked with arrows in the $\delta_1$ time series of Figs.~\ref{timeseries_r6e5} and \ref{timeseries_r2e7}.  In Figs.~\ref{profile_r6e5}(a,c,e) we illustrate the azimuthal profile of vertical velocity at $z=0.5$ and $r=0.48$ before and after the reorientations as blue and red curves respectively.   We calculate $\Delta \delta_1$, the change in the phase of the first Fourier mode, during the reorientations of the structures.  If the convective structure rotates by an angle $\Delta \delta_1$ during a reorientation, then, according to Eq.~(\ref{eq:fourier}), we can cancel the effects of this reorientation by subtracting $k\Delta \delta_1$ from the phases of all the positive $k$ modes, and by adding $k \Delta \delta_1$ to all the negative $k$ modes of the data recorded after the reorientation (Note that the $k=0$ mode is left unaltered).  We perform the above exercise on the Fourier modes of the data sets recorded after the specified reversals or reorientations.  Subsequently we compute the velocity field from the modified Fourier coefficients.  The modified velocity field, shown as green curves in Figs.~\ref{profile_r6e5}(b,d,f), matches quite well with the corresponding velocity profiles before the reorientations.  Thus we show that the convective structures essentially rotate by $\Delta \delta_1$ during the reorientations.  

The above features of our simulations are in general agreement with  the experimental results by \cite{cioni:1997},~\cite{brown:2005},~\cite{brown:2006}, and \cite{xi:2006}.  Note that the numbers of reorientations in the convection experiments are much larger than those observed in numerical simulations.    In the next subsections we analyze the detailed dynamics of the reorientation of the LSC using numerical data.

\subsection{Dynamics of reorientations}

Careful analysis of the phases and amplitudes of the Fourier modes provide important clues about the dynamics of the reorientations of LSC and the reversals of the vertical velocity.  The amplitudes of the Fourier modes vary significantly over time. We observe that during some of the reversals or reorientations,  the amplitude of the first Fourier mode almost vanishes; these reorientations are termed as ``cessation led''.  In the other reorientations that are termed as ``rotation led'' the Fourier modes continue to fluctuate around their mean values.  These kinds of reversals were reported by \cite{brown:2005}, \cite{brown:2006}, and \cite{xi:2006} based on their RBC experiment.  In the present paper we probe the dynamics of reversals using numerical simulations. 
\begin{figure}
  \begin{center}
 \includegraphics[width=0.78\columnwidth]{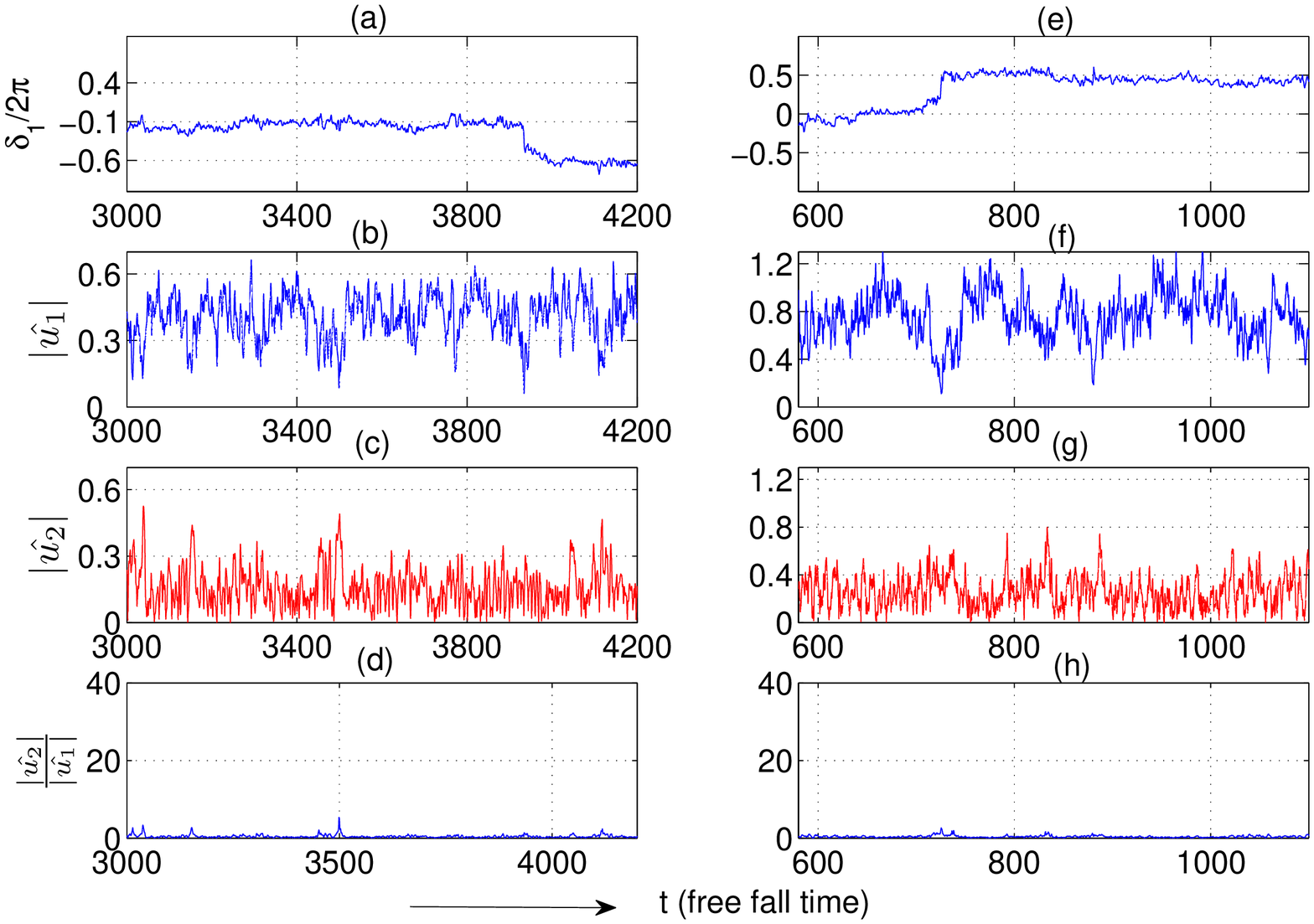}
  \end{center}
\caption{Time series of the phase of the first Fourier mode ($\delta_1$) (in units of $2\pi$), amplitude of the first Fourier mode  ($|\hat{u}_1|$), and amplitude of the second Fourier mode ($|\hat {u}_2|$ ) during the rotation-led reversals. (a), (b), (c), and (d) show the time series of $\delta_1$, $|\hat {u}_1|$, $|\hat {u}_2|$, and $|\hat {u}_2|/|\hat {u}_1|$ respectively for $R=6\times10^5$ on a ($33\times49\times97$) grid in which the reversal takes place near $t=3950$. (e), (f), (g), and (h) show the time series of $\delta_1$, $|\hat {u}_1|$, $|\hat {u}_2|$, and $|\hat {u}_2|/|\hat {u}_1|$ respectively for $R=2\times10^7$ on a ($100\times120\times201$) grid with a reorientation event near $t=700$.}
  \label{amp_ph_fullrev}
  \end{figure}
In Fig.~\ref{amp_ph_fullrev} we plot the phases and amplitudes of the first two Fourier modes during two rotation-led reorientations.  For $R=6\times10^5$ Fig.~\ref{amp_ph_fullrev}(a) exhibits the time series of the phase ($\delta_1/2\pi$, i.e., in units of $2\pi$).  Figs.~\ref{amp_ph_fullrev}(b,c) show the amplitudes of the first and second Fourier modes ($|\hat{u}_1|$, $|\hat{u}_2|$) respectively, while Fig.~\ref{amp_ph_fullrev}(d) shows their ratio $|\hat{u}_2|/|\hat{u}_1|$.   As shown in the figures, $\delta_1$ changes by approximately $\pi$ near $t \approx 3900$, but  $|\hat{u}_1|$ and $|\hat{u}_2|$ continue to fluctuate around their average values. However, $|\hat{u}_1|$ always dominates over $|\hat{u}_2|$, as evident from the plot of $|\hat{u}_2|/|\hat{u}_1|$ shown in Fig.~ \ref{amp_ph_fullrev}(d). Figures~\ref{amp_ph_fullrev}(e,f,g,h) exhibit similar features for $R=2\times10^7$.   Note however that $|\hat{u}_1|$ tends to have a small dip during the reorientation, but  $\Delta |\hat{u}_1|$   is much smaller compared to the cessation-led reversals to be described later. Note that both the above events lead to ``complete reversals" since $\Delta \delta_1 \approx \pi$.
\begin{figure}
 \begin{center}
\includegraphics[width=0.78\columnwidth]{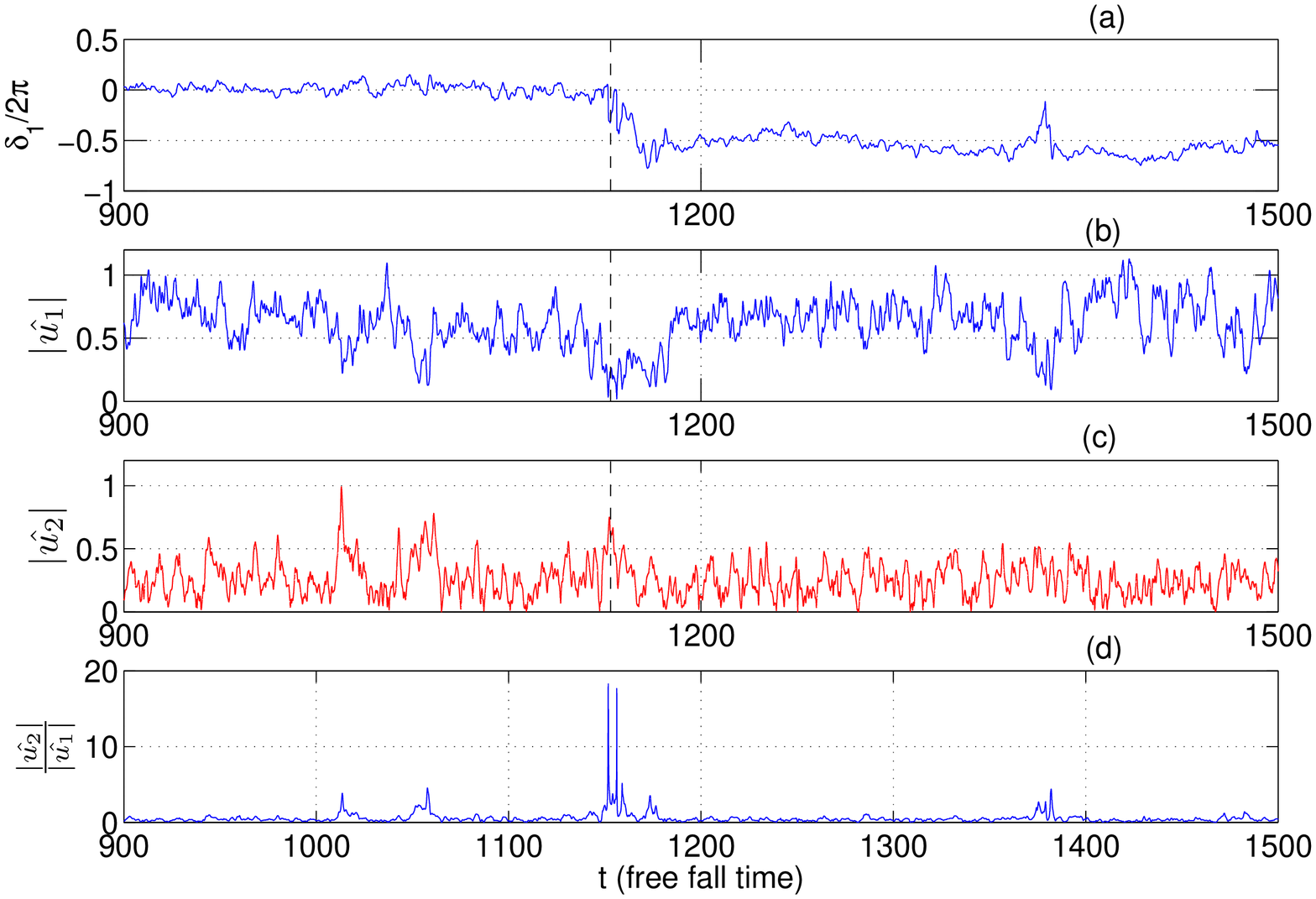}  
  \end{center}
  \caption{Time series of $\delta_1$,  $|\hat{u}_1|$, $|\hat{u}_2|$, and $|\hat{u}_2|/|\hat{u}_1|$ indicating cessation-led reorientation near $t\sim1175$ for $R=8\times10^6$ ($75\times96\times145$). The strength of $|\hat{u}_1|$ decreases significantly, while $|\hat{u}_2|/|\hat{u}_1|$ increases during this event.}
  \label{cess_rev_ra8e6}
  \end{figure}

In the other set of reversals or reorientations, we observe that the amplitude of the first Fourier mode ($|\hat{u}_1|$) tends to vanish during the event.  These set of reversals were termed as ``cessation led'' by~\cite{brown:2005}, \cite{brown:2006}, and \cite{xi:2006}.  We illustrate the cessation-led reversals using the following examples.

\begin{figure}
  \begin{center}
\includegraphics[width=0.78\columnwidth]{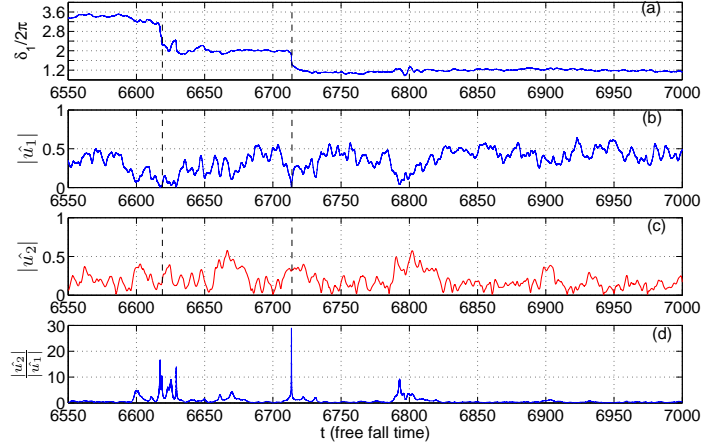}  
  \end{center}
  \caption{Time series of $\delta_1$,  $|\hat{u}_1|$, $|\hat{u}_2|$, and $|\hat{u}_2|/|\hat{u}_1|$  indicating cessation-led partial reorientations near $t\sim6625$ and $6710$ for $R=6\times10^5$ ($33\times49\times97$ grid).  For both the events $\Delta \delta_1$ are approximately $72^0$.} 
  \label{cess_partial_ra6e5}
  \end{figure}

Figures~\ref{cess_rev_ra8e6}(a) and~\ref{cess_rev_ra8e6}(b) exhibit the time series plots of  $\delta_1$ and $|\hat{u}_1|$ respectively for $R=8\times10^6$.   During the cessation at $t\sim1175$,  $\Delta\delta_1\approx \pi$ and  $|\hat{u}_1|$ tends to zero. The amplitude of the second Fourier mode  $|\hat{u}_2|$ however increases slightly above its average value during this event~(see Fig.~\ref{cess_rev_ra8e6}(c)). The fluctuations in $|\hat{u}_1|$ and $|\hat{u}_2|$ however tend to hide the above features.  Therefore we use $|\hat{u}_2| / |\hat{u}_1|$ to amplify the decrease in $|\hat{u}_1|$ and increase in $|\hat{u}_2|$ to be able to identify the cessation-led reversals clearly.  In ~\ref{cess_rev_ra8e6}(d) we plot  $|\hat{u}_2| / |\hat{u}_1|$ that exhibits a sharp peak during the event. This is an example of cessation-led complete reversal  since  $\Delta\delta_1\approx \pi$.  Figure~\ref{cess_partial_ra6e5} shows the corresponding quantities for another set of cessation-led reorientations for $R=6 \times 10^5$.    As evident from the time series, near $t\sim6620$,  $\Delta\delta_1\sim  (3.2-2)*2\pi \approx 72^0$.  Later at $t\sim6720$, $\Delta\delta_1\sim  (2-1.2)*2\pi \approx -72^0$.  These two reorientations of Fig.~\ref{cess_partial_ra6e5} are partial ones, and the LSC comes back to its original configuration after the two reorientations ($\Delta \delta_1  = (3.2-1.2)*2\pi$). During both these events $|\hat{u}_2| / |\hat{u}_1| $ has a sharp peak. 

\begin{figure}
  \begin{center}
\includegraphics[width=0.78\columnwidth]{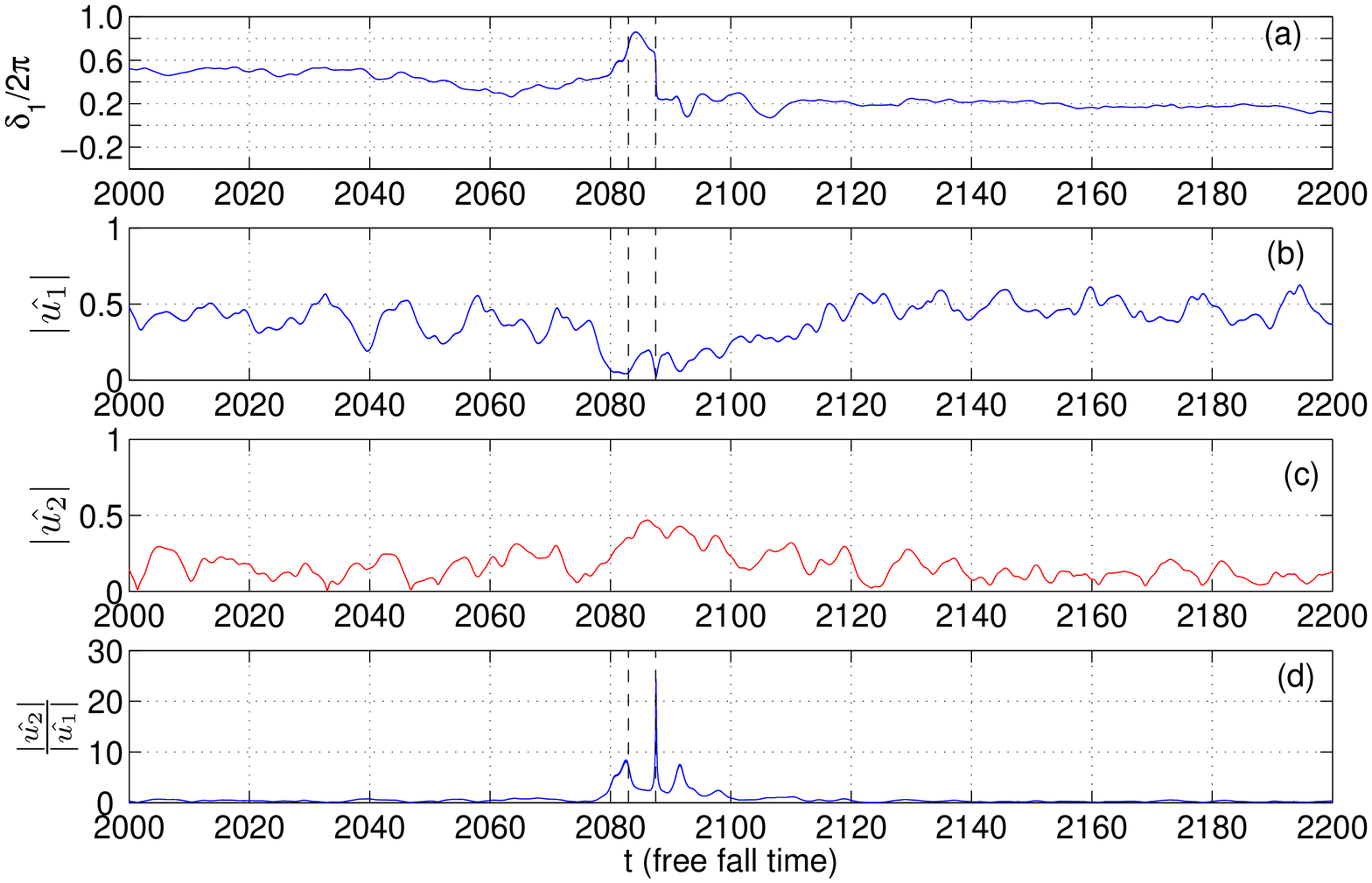}
  \end{center}
\caption{Time series plot of $\delta_1$, $|\hat{u}_1|$, $|\hat{u}_2|$, and $|\hat{u}_2|/|\hat{u}_1|$ indicating double cessation for $R=6\times10^5$ ($33\times49\times97$ grid). First cessation occurs at $t\sim2083$ and second one occurs at $t\sim2088$. During both the events $|\hat{u}_1|$ becomes weak and $|\hat{u}_2|$ dominates. (d) shows the two spikes in $|\hat{u}_2|/|\hat{u}_1|$ during the events.}
  \label{amp_ph_cessation1_r6e5}
  \end{figure}

\begin{figure}
  \begin{center}
\includegraphics[width=0.78\columnwidth]{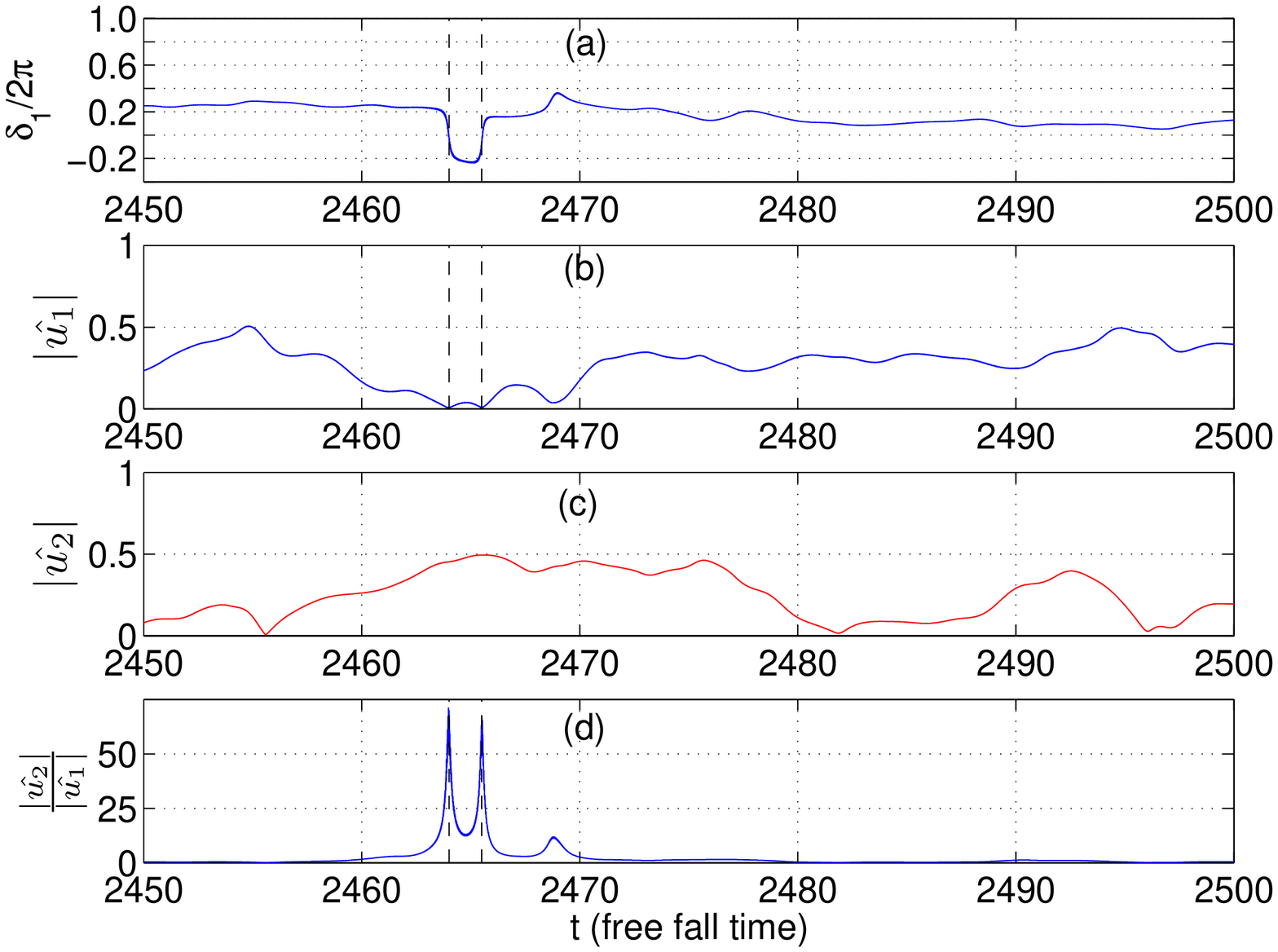}
  \end{center}
\caption{Time series plot of $\delta_1$, $|\hat{u}_1|$, $|\hat{u}_2|$, and $|\hat{u}_2|/|\hat{u}_1|$ indicating double cessation for $R=6\times10^5$ ($33\times49\times97$ grid). First cessation at $t\sim2464$ is followed by the second at $t\sim2466$. During both the events $|\hat{u}_1|$ becomes weak and $|\hat{u}_2|$ dominates. (d) shows the two spikes in $|\hat{u}_2|/|\hat{u}_1|$ during the events.}
  \label{amp_ph_cessation2_r6e5}
  \end{figure}

 In our simulations we also observe reorientations involving ``double cessations'', first observed by~\cite{xi:2006} in their experiments.   During these events, the phase $\delta_1$  changes in two stages in quick succession, first by $\theta_1$ and then second by $\theta_2$.  Therefore the net $\Delta \delta_1 \approx \theta_1 + \theta_2$. During double cessation, $|\hat{u}_1|$ vanishes  on two occasions separated by a small time gap (within an eddy turnover time).  $|\hat{u}_2|$ tends to increase during these times, hence  $|\hat{u}_2| / |\hat{u}_1| $ exhibits two peaks within a short time interval.  Two independent double cessation events have been  illustrated in Fig.~\ref{amp_ph_cessation1_r6e5}(a) and Fig.~\ref{amp_ph_cessation2_r6e5}(a) (for $R=6\times 10^5$). In Fig.~\ref{amp_ph_cessation1_r6e5}(a)  we observe the first cessation at $t\approx 2083$ followed by the second cessation at $t\approx 2088$, and $\theta_1 \approx -\theta_2 \approx \pi$.  In Fig.~\ref{amp_ph_cessation2_r6e5}(a) the first cessation occurs at $t \approx 2464$ followed by the second cessation at $t \approx 2466$ with $\theta_1 \approx -\theta_2 \approx 0.8\pi$.   Since the net change in the phase $\delta_1$ is approximately zero, the final configuration of LSC is similar to its original configuration.   Note that one eddy turn time for the corresponding run is approximately 26 free fall time (see Table 2).  Hence the two double cessations occur within $0.2$ and $0.08$ eddy turnover time respectively, which is consistent with the experimental result of~\cite{xi:2006}.

\begin{figure}
  \begin{center}
\includegraphics[width=0.78\columnwidth]{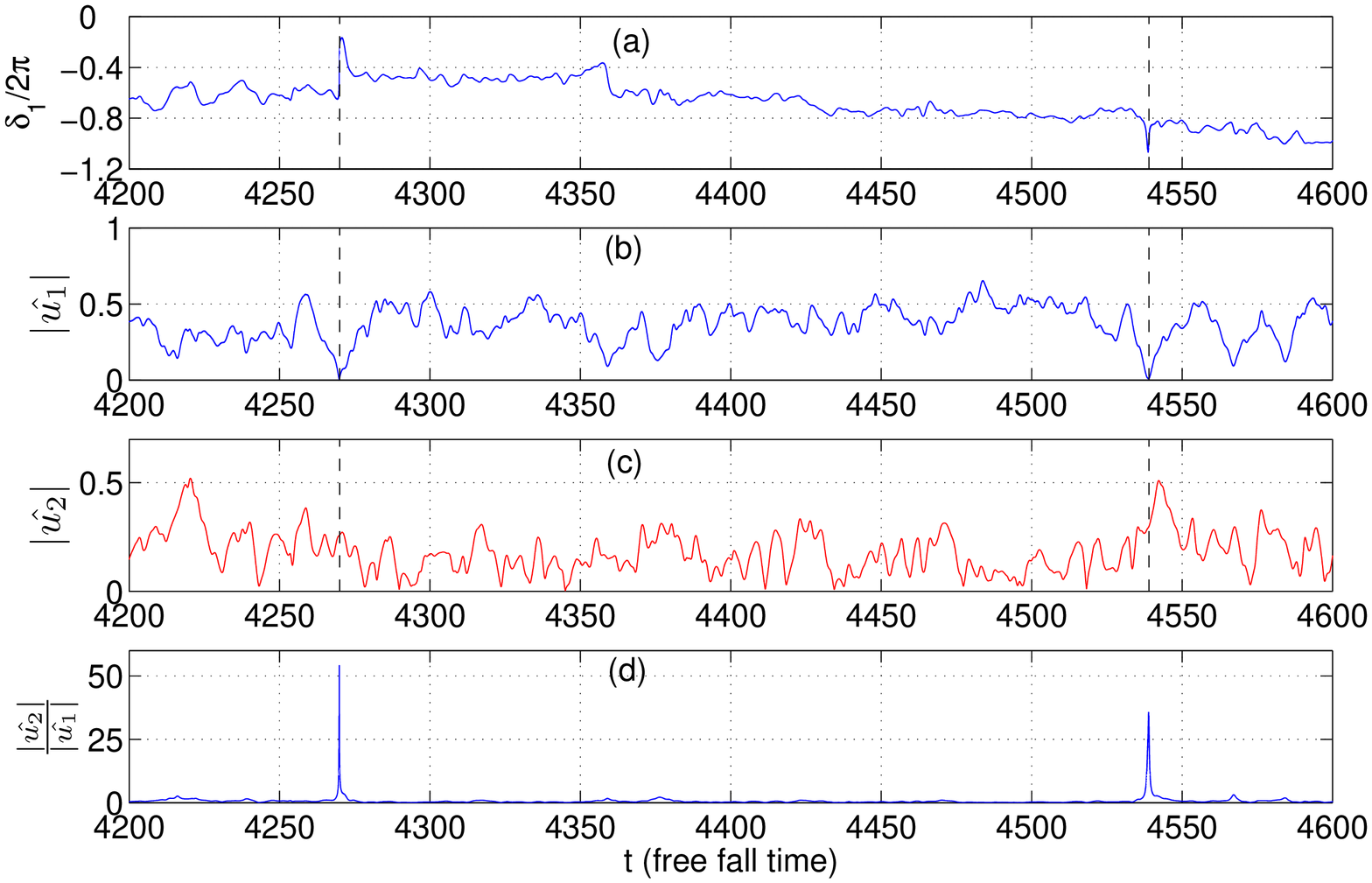}
  \end{center}
\caption{ For $R=6\times10^5$  time series plot of $\delta_1$, $|\hat{u}_1|$, $|\hat{u}_2|$, and $|\hat{u}_2|/|\hat{u}_1|$ depicting events similar to double  cessations ($33\times49\times97$ grid).  $\Delta\delta_1\sim72^0$ near $t\sim4270$, while $\Delta\delta_1\approx 0$ near $t\sim4540$. During both the events $|\hat{u}_1| \rightarrow 0$ but the ratio $|\hat{u}_2|/|\hat{u}_1|$ increases significantly.}
  \label{amp_ph_cess_r6e5}
  \end{figure}
 For $R= 6 \times 10^5$ we observe two events that are similar to the above mentioned double cessation events (see Fig.~\ref{amp_ph_cess_r6e5}(a,b,c,d)).  Near $t \approx 4540$, $\delta_1$ changes twice, $\theta_1 \approx 0.6\pi$ and $ \theta_2 \approx -0.6 \pi$ with net $\Delta \delta_1 \approx 0$.  These two events however occur very close to each other, and we observe only one peak peak for $|\hat{u}_2| / |\hat{u}_1| $, rather than any double peaks indicative of double cessation.   Another cessation-led reorientation occurs near $t\approx 4270$ that resembles the above double-cessation event except one major difference.  Here $\theta_1 \approx 0.8\pi$ and $ \theta_2 \approx -0.6 \pi$, hence LSC reorients by a net angle after the event.   Note that for the double cessation reported by \cite{xi:2006}  $\theta_1 \approx -\theta_2$ and the net change in $\delta_1$ is zero.    In our numerical simulations we find such events.  In addition we also observe a double cessation where $\theta_1 \ne -\theta_2$, and the LSC reorients by a finite angle after the event.
\begin{figure}
  \begin{center}
   \includegraphics[width=0.78\columnwidth]{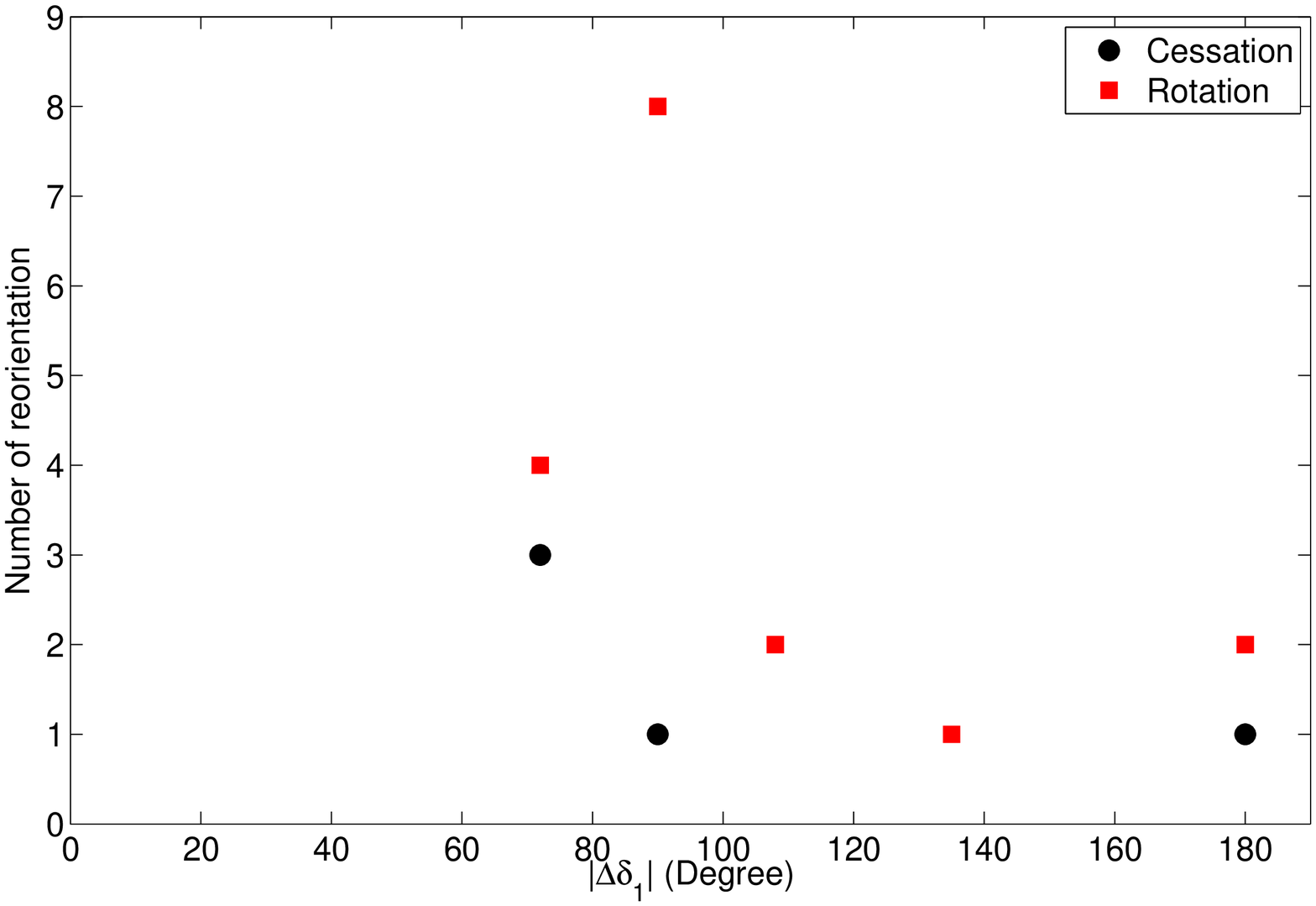}
  \end{center}
  \caption{The distribution of the change in the phase of the first Fourier mode ($|\Delta \delta_1 |$) of LSC during reorientations.  Criteria of  \cite{brown:2005} and \cite{brown:2006} have been used for identifying reorientations of LSC. }
  \label{rotation_reorientation}
  \end{figure}
Figure~\ref{rotation_reorientation} illustrates the distribution of  $\Delta \delta_1$ of LSC observed in our numerical simulations.   We observe three complete reorientations ($\Delta \delta_1 \approx \pi$) among which two are rotation-led while one of them is cessation-led.   The number of  observed partial reorientations are rather large (total 19).  Among the partial reorientation events, 15 of them are rotation-led, while 4 of them are cessation-led.   We also observe 4 double cessations among whom three of them have $\Delta \delta_1 \approx 0$ while the fourth one has $\Delta \delta_1 \approx 0.2 \pi$. The above observations indicate that cessation-led events are rarer compared to the rotation-led ones, an observation consistent with those of \cite {brown:2005} and \cite {brown:2006}.   The number of reorientations observed in our simulations are far fewer compared to those observed in the experiments of \cite {brown:2005}, \cite{brown:2006}, \cite{cioni:1997}, and ~\cite{xi:2006}.  Consequently we are unable to perform statistics of reorientations similar to those by \cite{brown:2005} and \cite {brown:2006}. 
\begin{figure}
  \begin{center}
   \includegraphics[height=!,width=15cm]{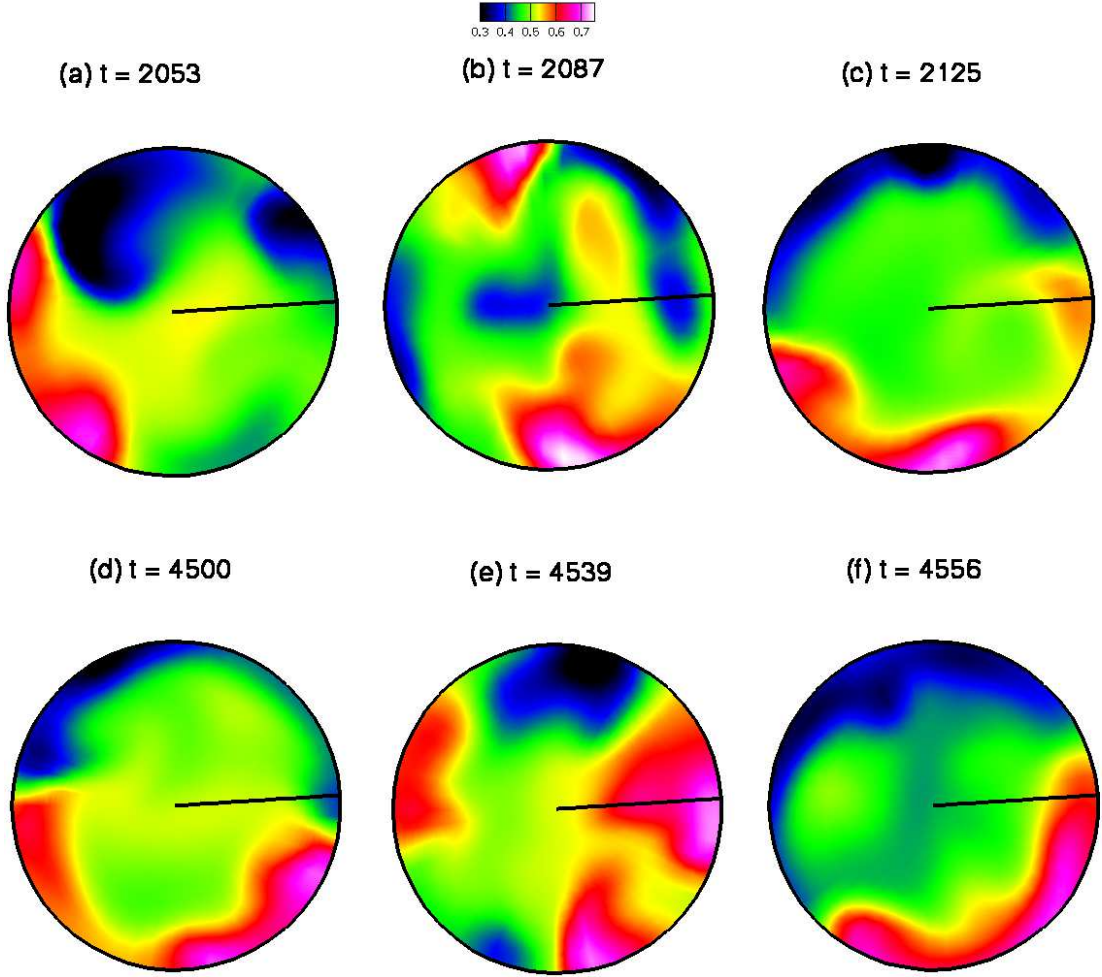}
  \end{center}
\caption{Temperature snapshots at the mid plane ($z=0.5$) during cessation-led reorientations near $t\sim2087$  and $4539$ for $R=6\times10^5$.  Dipolar structures are dominant before and after the reorientations, but quadrupolar structure is prominent during the reorientations (at $t=2087$ and $4539$).  }
  \label{snap_cess_r6e5}
  \end{figure}

 In our numerical simulations we observe that the higher Fourier modes play interesting role during reorientations.  These kind of investigations were missing in earlier experimental works.  We find that during the cessation-led reorientations $|\hat{u}_2|$ tends to become significant  when the  $|\hat{u}_1|$ tends to vanish. This feature is evident from the significant increase in the magnitude of $|\hat{u}_2|/|\hat{u}_1|$ during the cessation-led reorientations .~\cite{brown:2005} and \cite{brown:2006}  interpret such events as cessation of the circulation, followed by a restart in a randomly chosen new direction. Our simulation results are consistent with the above interpretation, yet another scenario is also possible.  During the cessation, the circulation structure corresponding to $\hat{u}_1$ (dipolar in a horizontal plane) becomes very weak, and the quadrupole structure corresponding to $\hat{u}_2$  becomes dominant.  After the cessation, the quadrupole structure disappears and dipolar structure reappears with a shift in the azimuthal direction.  These features have been illustrated in Fig.~\ref{snap_cess_r6e5} in which we show the horizontal profile of temperature in $z=0.5$  plane before, during, and after two cessation-led reorientations for $R=6\times10^5$.  Figures~\ref{snap_cess_r6e5}(a,b,c) represent the temperature snapshots near a double cessation event (near $t=2100$ of Fig.~\ref{amp_ph_cessation1_r6e5}).  The system starts from $\hat{u}_1$ dominant state with approximate profile as $\cos(\theta+\delta_1$) of Fig.~\ref{snap_cess_r6e5}(a).  During the cessation, the system profile appears as Fig.~\ref{snap_cess_r6e5}(b) which has $\hat{u}_2$ as the most prominent Fourier mode with profile as $\cos(2 \theta+\delta_2)$.  After the double cessation, the system return to $\hat{u}_1$ dominant structure as evident from Fig.~\ref{snap_cess_r6e5}(c).  Similar features are observed for another cessation-led reorientation near $t \approx 4500$ for $R=6\times10^5$  shown in Figs.~\ref{snap_cess_r6e5}(d,e,f).   Note that in the rotation-led reorientations, the dipolar structure continues to be dominant during the reorientation itself.

In this section we present some of the complex dynamics of the reorientations of LSC.  The reversals of the vertical velocity have been shown to be intimately related to the reorientations of LSC.   We observe rotation-led and cessation-led reorientations, which were first reported by \cite{brown:2005} and \cite{brown:2006}, and observed later in other experiments \cite[]{xi:2006,xi:2007,xi:2008}.  In our simulations  we also find double cessation, previously observed by~\cite{xi:2006}.  A new feature of our simulation is a double cessation in which the orientation of the LSC after the event is different from the original orientation.   In our analysis we study higher Fourier modes, notably $\hat{u}_2$, which was missing in earlier work on reversals.  We propose a new interpretation of the cessation-led reorientations, and argue that the LSC transforms from dipolar ($\hat{u}_1$ dominant) to quadrupolar ($\hat{u}_2$ dominant), and then back to dipolar ($\hat{u}_1$ dominant) structure during this event.

\section{\label{sec5}Summary and Conclusions}
In summary, we numerically compute  turbulent convective flows for a cylindrical geometry and study the characteristics of the LSC, in particular those related to flow reversals.   The first Fourier mode along the azimuthal direction is the most dominant mode; the phase of this mode ($\delta_1$) is used as a measure of the orientation of  LSC. Our numerical results  and earlier experimental results indicate that the LSC occasionally reorients itself by rotating along the azimuthal direction by an arbitrary angle.   When the reorientation angle is around $\pi$, the vertical velocity at {\it all} probes change sign signaling complete reversal of the flow.  A partial reversal is observed when the reorientation angle is less than $\pi$.  We therefore argue that the primary cause of the flow reversal is the azimuthal reorientation of the LSC. 

In our numerical simulations we observe two kinds of reorientations: (a) rotation-led, and  (b) cessation-led, earlier observed by~\cite{brown:2005}, \cite{brown:2006}, and \cite{xi:2006}. The rotation-led reversals involve rotation of the LSC without appreciable reduction in circulation strength, i.e., the amplitude of $|\hat{u}_1|$ remains finite.  In the cessation-led reversals, $|\hat{u}_1| \rightarrow 0$, which is interpreted by \cite{brown:2005} and \cite{brown:2006} as cessation of circulation followed by a restart in a randomly chosen new direction.  During this event, our numerical simulations reveal that the ratio of the amplitude of the second Fourier mode and the first Fourier mode,  i.e., $|\hat{u}_2|/|\hat{u}_1|$,  increases significantly, and then it comes back to its original level.  These properties of the Fourier modes reveal that during the cessation, the LSC transforms from a dipolar-like structure to a quadrupolar-like structure, and then back to a dipolar-like structure. These features appear to have certain similarities with the reversals of magnetic field in dynamo~\cite[]{reversal_dipole}. The role of higher Fourier modes for reversals and reorientations of LSC have been highlighted in our simulations for the first time.

We also observe double cessation in our simulations.   \cite{xi:2006} had observed in their experiments that the LSC returns to its original orientation (approximately) after a double cessation.  We however find  in our numerical simulations that the change of the orientations of LSC during some double cessations are zero, while in some others, the changes are nonzero.

Our numerical simulations reproduced many features (``rotation-led reorientations", ``cessation-led reorientations'', ``double cessation'', etc.) observed  in convection experiments by \cite{cioni:1997}, \cite{brown:2005}, \cite{brown:2006}, \cite{xi:2006},  and \cite{niemela:2001} for Rayleigh numbers greater than those in our simulations. 
Several researchers have argued that the reorientations, including reversals, of LSC occur only  under strong turbulence regimes (say $R > 10^8$).  We however observe that the nature of convective flows and reorientations are very similar for the range of Rayleigh numbers studied by us ($R=6\times10^5$ to $R=3\times10^7$).  The lowest Rayleigh number $R=6\times 10^5$ is probably in the weak turbulence regime where the LSC is not well organized, and   $R=3\times 10^7$ is in the lower end of strong turbulence regime.   We could not carry out simulations for even larger Rayleigh numbers due to very expensive computational requirements for these runs.   However we believe that reversals in high Rayleigh number regimes are highly likely to be similar to those presented in the simulations.

{\bf Acknowledgements:} 
We thank Krishna Kumar, St\'{e}phan Fauve, Pankaj Wahi,  Supriyo Paul, and Pinaki Pal for very useful discussions and comments.   We also thank the reviewers of this paper who provided many useful comments, interesting ideas and suggestions that have contributed significantly to the improvement of the paper.  This work was supported by a research grant by the Department of Science and Technology, India as Swarnajayanti fellowship to MKV.

\label{lastpage}
\end{document}